# Spectral Classification of Galaxies: An Orthogonal Approach

A. J. Connolly and A. S. Szalay[1]

Department of Physics and Astronomy, The Johns Hopkins University, Baltimore, MD 21218
Electronic mail: ajc@skysrv.pha.jhu.edu

M. A. Bershady[2,3]

Department of Astronomy & Astrophysics, Pennsylvania State University, University Park,
PA 16803

A. L. Kinney and D. Calzetti

Space Telescope Science Institute, Baltimore MD 21218

## ABSTRACT

Classification of galaxy spectral energy distributions in terms of orthogonal basis functions provides an objective means of estimating the number of significant spectral components that comprise a particular galaxy type. We apply the Karhunen-Loève transform to derive a spectral eigensystem from a sample of ten galaxy spectral energy distributions. These spectra cover a wavelength range of 1200 Å to 1 $\mu m$ and galaxy morphologies from elliptical to starburst. We find that the distribution of spectral types can be fully described by the first two eigenvectors (or eigenspectra). The derived eigenbasis is affected by the normalization of the original spectral energy distributions. We investigate different normalization and weighting schemes, including weighting to the same bolometric magnitude and weighting by the observed distribution of morphological types. Projecting the spectral energy distributions on to their eigenspectra we find that the coefficients define a simple spectral classification scheme. The galaxy spectral types can then be described in terms of a one parameter family (the angle in the plane of the first two eigenvectors). We find a strong correlation in the mean between our spectral classifications and those determined from published morphological classifications.

*Subject headings:* galaxies, spectroscopy, classifying, mathematical methods

---

[1]Department of Physics, Eötvös University, Budapest, Hungary, H-1088

[2]Lick Observatory, University of California Santa Cruz, CA 95064

[3]Hubble Fellow

# 1. INTRODUCTION

Classification of galaxy spectra is hindered by the multivariate nature of the properties of galaxies. Determining which spectral components are of significance for classification, by pairwise correlation of the measurable attributes (e.g., equivalent widths of emission lines or continuum features) may lead to confusing results as many of the properties are not independent. We can, however, determine the minimum number of dimensions that are required to describe the multidimensional distribution of variables by applying a principal component analysis (PCA; Chatfield & Collins 1980). This assumes that a linear relation exists between all variables and is simply a rotation of the axes within the multidimensional space to a system where the new axes are orthogonal and aligned with the directions of maximal variance.

PCA has been applied to the problem of spectral classification; from using extracted features (Boroson & Green 1992) to using the spectral data directly (Francis *et al.* 1993). For QSO spectra it was found that, to account for their continuum and emission line properties, over ten principal components were required. However, using only three components broad absorption line QSO's could be separated from the full QSO sample (Francis *et al.* 1993).

This paper describes the application of an eigensystem or PCA analysis to the classification of optical and UV galaxy spectra. We apply a variant of the PCA technique for deriving the eigenvectors, the Karhunen-Loève transform. We aim to determine the optimal combination of spectral components required to "build" a galaxy spectrum and apply this to the classification of galaxies ranging from blue starbursts to red ellipticals. The advantages of this approach over standard PCA are discussed.

In §2 we describe the spectral data used in our analysis. The Karhunen-Loève transform as applied to these data is outlined in §3 together with a discussion on the effect of different spectral normalizations. We decompose the SED's into linear combinations of their eigenspectra and use these coefficients as a spectral classification scheme in §4.

# 2. DATA

The data used in the present analysis are the template UV–optical spectra of the central regions of quiescent and star-forming galaxies discussed by Kinney *et al.* (1994) and by Calzetti *et al.* (1994). We give here a brief description of the dataset.

The sample from which the templates were derived includes the spectra of 31 quiescent galaxies of different morphological types, from E to Sc, and the spectra of 39 starburst galaxies, mainly Irregulars and galaxies with disturbed morphologies. The main selection criterion for the galaxies included in the sample is their detectability at UV wavelengths; therefore, the sample of quiescent galaxies, which are typically UV faint, includes only bright, local galaxies, while the sample of starbursts, which have a large range in UV brightness, includes galaxies up to a redshift of z∼0.03.

The UV spectra are archival IUE spectra collected in the Atlas of Kinney *et al.* (1993) and the Atlas of Koratkar *et al.* (1994), or are from original IUE observations as part of a program to produce templates of normal galaxies (Kinney *et al.* 1994). The UV spectra cover the wavelength range 1200-3200 Å, with a resolution of ∼ 6 Å. The optical spectra were obtained at the KPNO and CTIO observing facilities (McQuade *et al.* 1994, Storchi-Bergmann *et al.* 1994). The CTIO spectra were obtained with a long slit of $10''$ width; subsequently a window $20''$ long was extracted to match the $10'' \times 20''$ IUE aperture. The CTIO spectra cover the wavelength range 3200-10000 Å with a resolution of ∼ 8 Å. The KPNO spectra were obtained using a circular aperture of $13.5''$ diameter. The difference in flux due to the difference in size between the optical and the IUE apertures is at most 30%, which is within the uncertainty of the absolute optical fluxes. The KPNO spectra cover the wavelength range 3200-7700Å with a resolution of 10Å.

The UV–optical spectra of the quiescent galaxies were grouped according to morphological type: E, S0, Sa, Sb, and Sc, respectively (see Table 2 in Kinney *et al.* 1994). The spectra of starburst galaxies were grouped according to increasing values of the color excess E(B-V): from S1, with E(B-V)=0.05, to S6, with E(B-V)=0.70 (see Table 3 in Calzetti *et al.*, 1994).

Prior to analysis, all the spectra were shifted back to the rest frame and were corrected for the foreground extinction of the Milky Way. Within each starburst group (S1 to S6), the UV–optical spectra were rescaled to a common flux value and were averaged after weighing each spectrum by its signal-to-noise ratio, to produce the final template. The templates of quiescent galaxies were constructed by analyzing and averaging separately the UV and the optical spectra, owing to the fact that optical spectra were not obtained for all galaxies observed in the UV. Within each morphological group (E to Sc), the optical spectra were averaged after rescaling the fluxes to a common value and weighting each spectrum by the exposure time (analogous to a S/N weight, since the optical spectra of the quiescent galaxies do not show a large range in flux level). The UV spectra of the quiescent galaxies were averaged according to the same procedure adopted for the starburst galaxies.

We exclude the composite Sc galaxy spectrum from our analysis. This is due to the poor signal-to-noise of the optical and UV data used in its construction (see Kinney *et al.*, 1994). In order to emphasize the emission from the stellar continuum in the templates, the strongest nebular emission lines in the starbursts were removed by fitting the continuum at the base of each line with a first order polynomial.

## 3.  AN EIGENSYSTEM ANALYSIS OF GALAXY SPECTRA

### 3.1.  Derivation of eigenspectra

Much has been written about the use of Principal Component Analyses in studies of the multivariate distribution of astronomical data (Efstathiou & Fall 1984). We limit ourselves, therefore, to a discussion of the Karhunen-Loève transform (KL; Karhunen 1947, Loève 1948) and its application to a set of spectral energy distributions. Each SED can be thought of as an axis within a multidimensional hyperspace, $f_{\lambda i}$, where $\lambda$ is the wavelength and $i$ the spectral type. From this hyperspace we wish to construct normal spectra, $e_{\lambda i}$, that represent an orthogonal basis or eigensystem. We will call these new spectra, eigenspectra.

From the set of ten SED's we construct a $10 \times 10$ symmetric correlation matrix, $C_{ij}$, such that,

$$C_{ij} = \frac{f_i.f_j}{N_i N_j} = \frac{1}{N_i N_j} \sum_\lambda f_{\lambda i} f_{\lambda j} W_\lambda. \tag{1}$$

Each element of $C_{ij}$ is the scalar product of the corresponding SED's, divided by a normalization $N_i$. The normalization schemes adopted in this paper are discussed in §3.2. $W_\lambda$ can weight the different features within the spectra (i.e. we can emphasize particular absorption features or remove the effect of emission lines within our analysis). We adopt a uniform weight, in wavelength space, for this paper.

To determine the eigenbasis, $e_{\lambda i}$, we require a matrix, $R$, that reduces the correlation matrix to its diagonal form, $\Lambda$, i.e.,

$$R^\dagger C R = \Lambda. \tag{2}$$

The diagonal components of $\Lambda$ are the eigenvalues of the system. The corresponding eigenbasis is, therefore,

$$e'_{\lambda i} = R^\dagger f_{\lambda i} \tag{3}$$

and satisfies the property

$$e'_i.e'^\dagger_j = \gamma_i \delta_{ij} \tag{4}$$

with $\gamma_i$ the eigenvalue of each eigenspectrum. Normalizing these vectors we construct the orthonormal eigenbasis, $e_{\lambda i}$ which will be used below. This analysis is known as the Karhunen-Loève transform and in its discrete form is equivalent to a principal component analysis.

It is clear that each SED can be represented, with no error, by a linear combination of the eigenspectra, i.e.,

$$f_{\lambda i} = \sum_{j=1}^m y_{ij} e_{\lambda j}, \tag{5}$$

where $m$ is the total number of eigenspectra and $y_{ij}$ is the weight of the $j$th eigenspectrum in the $i$th galaxy. However, to reduce the dimensionality of the distribution of data we wish to know how well we can describe each spectral type using a subset of these eigenspectra. We, therefore, define a mean square error for the whole sample as,

$$\langle \epsilon^2 \rangle = \sum_{i=1}^{m} \langle \epsilon_i^2 \rangle, \tag{6}$$

$$= \sum_{i=1}^{m} [f_{\lambda i} - \sum_{j=1}^{n} y_{ij} e_{\lambda j}]^2, \tag{7}$$

where $n < m$. The minimum mean square error is related to the eigenvalues by the following simple relation,

$$\langle \epsilon^2 \rangle = \sum_{j=n+1}^{m} \gamma_j, \tag{8}$$

if $\gamma_i > \gamma_{i+1}$. Thus the eigenvalues represent the statistical relevance of the corresponding eigenspectra in the galaxy sample: they describe the relative power of each eigenspectrum – small eigenvalues mean irrelevant contributions. The above truncation of this series expansion (i.e. $n < m$) provides an optimal filtering (in a quadratic sense) of the relevant spectral information from the noise. The dimensionality of the data can be inferred from the relative magnitudes of the eigenvalues (see Tables 1, 2 and 3).

For our analysis of spectral eigenfunctions the KL transform offers several advantages over a simple PCA. While both essentially derive the moments of the distribution of data points they do so using different projections of the data. In the PCA the diagonalization of the matrix would be performed on a symmetric matrix with size the number of spectral elements (2207), but only of a rank ten, i.e. the number of independent components in each spectral resolution element. By projecting the distribution on to the axes defined by the different spectral types we reduce this problem to a 10×10 matrix diagonalization. Secondly, using the KL approach it is straightforward to introduce different weighting schemes to the analysis to emphasize different wavelength regions. For problems where the number of variables is greater than the number of data points the KL transform clearly offers a more elegant and efficient solution.

## 3.2. Normalization of galaxy spectra

The results of the KL transform are, in principle, sensitive to the normalization of the original SED's (see Eqn 1). We show, however, that all schemes with sensible physical motivations give identical, robust results. The normalization dependence can be understood by considering the features present in the spectra (e.g. the absorption and emission lines, excess of blue light). Each type of galaxy contains different sets of these features. By preferentially weighting particular types of galaxies we can emphasize some features and suppress others, altering the resultant eigenspectra. This is analogous to the rescaling of the axes in a PCA (e.g. to unit variance).

The ten galaxy spectra represent ten vectors in a 2207 dimensional space. Since the absolute fluxes are not relevant here, each of these vectors only indicates a direction. The most natural normalization is, therefore, for all of them to have the same unit length: the scalar product (as defined in Eqn 2) is set to unity. Alternatively one can argue for using the total flux as a normalization. In the multidimensional representation this will result in a slight change of the lengths of the ten vectors. Since the overall spectral envelopes are not grossly different, these changes will be small. Our results will prove this point quantitatively.

If we change the lengths of some of the ten vectors drastically, then the results may change as well: e.g. if all vectors lie in a plane, evenly distributed, but we shrink them by a large factor in one direction, within the plane, then we create a one-dimensional distribution by this massive rescaling. Our third normalization scheme, correcting the norm with the comoving distribution of morphological types observed locally, performs this very operation – effectively the galaxies with star-formation are removed from the sample – and we get a result in accordance.

In the following section we investigate the effect of these three normalization schemes quantitatively: normalization by the scalar product, normalization by the total flux and correcting by the observed distribution of morphological types. Normalization of the spectra changes the sum of the eigenvalues as well as the eigenspectra. For comparative purposes, therefore, we include the relative power carried by each eigenspectrum in Tables 1, 2 and 3.

### 3.2.1. Normalization by the scalar product: a uniform prior

The results of the KL transformation for the scalar product normalization, $N_i^S = \sqrt{\sum_\lambda f_{\lambda i}^2 W_\lambda}$, are shown in Fig 2. The eigenspectra are ordered by decreasing eigenvalues. The eigenvalue of each eigenspectrum is given in Column 1 of Table 1. The power of that eigenspectra (defined as the eigenvalue divided by the sum of the eigenvalues) is given in Column 2 and the relative contributions of the SED's to an eigenspectrum are given in Columns 3 through 12.

Over 98.5% of the power in the correlation matrix is contributed by the first two eigenspectra. The first component represents the "typical galaxy" spectrum and is essentially the mean of the ten SED's, modulated by the mean square of their integrated fluxes. This modulation gives a lower weight to the extremes of the spectral types, with the red starburst galaxies (S4, S5, S6) having $\sim 15\%$ more weight than either the blue starbursts or early type galaxies. The second eigenspectrum shows the "typical deviation" from the mean galaxy. In Fig 2 we show that the red and blue ends of the spectrum are anticorrelated, with a zero crossing at $\sim 4400$ Å.

By the third eigenspectrum the power contained in that component has fallen to 0.6%. The signal in this mode comes principally from the S6 starburst. This excess signal implies that the spectral energy distribution of S6 deviates from the simple obscuration sequence defined by the starburst templates. Indeed, the S6 template was constructed by including, among others, three spectra in excess of the mean obscuration trend defined in Calzetti *et al.* (1994, see their Figs. 5a and 18). The flux below 5000 Å may be evidence for a residual population of late B to F stars, in the central regions of these galaxies, from previous bursts of star formation. We will refer to this component as post starburst activity in the subsequent discussion.

Subsequent eigenspectra contribute less than 0.5% of the power in the correlation matrix. From their spectra in Fig 2 it is clear that they contain almost no signal. There do remain residual features within the spectra, both real and artifacts of the data reduction. For example, in the fifth component there is the presence of a residual sky line at 5577 Å and a MgI absorption feature at 5175 Å. In the eighth eigenspectrum the uncertainty in matching the UV and optical spectra appears as a feature at 3300 Å. That these residual features are already apparent by the fifth eigenspectrum is further indication that the overall continuum can be described by just the first few eigenspectra.

### 3.2.2. Normalization by the integrated flux

A more physically motivated normalization is to normalize the SED spectra to the same total flux, $N_i^F = \sum_\lambda f_{\lambda i}$. This is equivalent to normalizing to the same bolometric magnitude and allows an additional parameter in the classification, i.e. an absolute magnitude.

The derived eigenvalues and weights are given in Table 2. The ordering of the eigenmodes are as given previously. Qualitatively the first five eigenspectra are the same as those derived using the scalar product normalization. If, however, we consider the relative weighting of each of the SED's then we see that, for the first component, the weights decrease monotonically from blue to red. The resultant spectrum is, therefore, bluer than that derived above. Quantitatively the difference between the two normalizations occurs at less than the 1% level, for the first three eigenspectra.

The differences between these results and those from the scalar product normalization simply reflect the fact that the mean square of the flux and the mean of the flux are not identical. That the two sets of eigenspectra differ at only the 1% level shows that the differences between these measures (mean square and mean) are minor and do not affect the dimensionality of the distribution (i.e. the shapes of the SED's can be described by continuous functions). This may not be altogether surprising as we have removed much of the discontinuities by masking out the strong emission lines.

As the results of the scalar product normalization and flux normalization are similar we will only consider the scalar product in our subsequent analysis. We do this because normalizing by the scalar product produces normal vectors which offer distinct advantages for the classification discussed below.

### 3.2.3. Morphological type normalization

The previous two normalizations take no account of the comoving distribution of morphological types, all SED's being counted equally. This is essentially a stratified sampling technique, where we sample in equal spacing based on the limits of a distribution rather than sampling in steps of equal number (i.e. histogram equalization). We know, however, that the distribution of SED's does not match the observed distribution of spectral types. Galaxies undergoing intense bursts of star formation represent less than 5% of the total number of local galaxies while in our sample they comprise over 50% of the SED sample.

To account for this we renormalize the correlation matrix derived in §3.2.2 by the relative morphological populations, $p_i$, of the SED's, i.e. $N_i^M = N_i^S \sqrt{p_i m}$. The distribution of morphological types is taken from King & Ellis (1985). As the resolution of spectral types is poor, with no information on the distribution of starburst galaxies, we assume that starburst galaxies account for 3% of the local population of galaxies and make no attempt to distinguish between different types of starburst spectra.

The weights for all eigenspectra are given in Table 3 and the first two eigenspectra are shown in Fig 3. The ordering of the eigenspectra and the columns of the table are as previously stated. The most obvious difference from the previous two analyses is that the number of significant components has been reduced from two to one. The first component accounts for over 97% of the power within the correlation matrix. The second component is reduced to less than 2%. This is consistent with our previous interpretation. We have seen that the star formation is a principal axis within the 2207 dimensional space. By compressing the axis corresponding to the star formation (i.e. the renormalization of the starburst galaxies) we remove this axis in our results. We are left with a single spectral component being able to describe the early and late type galaxies.

## 4. CLASSIFICATION FROM EIGENSPECTRA

### 4.1. Decomposition of the galaxy spectra

We can determine the relative contribution of each eigenspectrum to a particular SED by calculating the respective expansion coefficients, $y_{ij}$, from the scalar product of the eigenspectrum with the normalized SED. We derive these coefficients for the eigenspectra normalized by the scalar product and those renormalized by the distribution of morphological types. The coefficients for each SED are given in Tables 4 and 5.

As seen in §3.2 the first two eigenspectra dominate the coefficients for the scalar product normalization, the sum of the first two eigenvalues carrying 99% of the total power. To demonstrate this, we reconstruct the SED's from just these two eigenspectra. This amounts to an optimal filtering of the spectral data with $n = 2$ (see Eqn 6). We show the resulting spectra and the residuals, for the scalar product normalization, in Fig 4. All ten spectra are quite well reproduced and the residuals are negligible, with the exception of galaxy S6 where, as noted previously, there remain residuals consistent with post starburst activity. To quantify this we derive the relative error, i.e. the root mean square error divided by the mean flux, for each reconstruction. This statistic is shown for the scalar product and morphological normalizations (with $n = 2$ and $n = 3$) in Fig 5.

The qualitative results shown in Fig 4 are borne out by the measures of the relative error. For the scalar product normalization, with $n = 2$, the S6 and S1 SED's have the largest relative error. This is due to the presence of a post starburst phase (as shown by the third eigenspectrum). By including the third component the relative error for S6 and S1 drops by a factor of two. Using the normalization by the distribution of morphological types the starburst galaxies are more poorly represented by the first two eigenvalues (with errors typically three times larger than those of early type galaxies). This is hardly a surprise, as they were given a much lower weight so that none of their representative features could make it into the first two eigenspectra. Including the third component reduces the relative error for the starburst galaxies but they remain poorly constrained.

## 4.2. Dimensionality and classification

In the previous section we show how well the first two eigenspectra represent all ten SED's in the sample. We note that the relative weights of the expansion coefficients in Tables 4 and 5 form a monotonic progression as a function of spectral type. These values can, therefore, be used as a way of classifying the galaxies. The first eigenspectrum gives the mean spectral type and the second eigenspectrum the deviation from this "typical galaxy". The classification of a particular is dependent on the modulation of the amount of red vs. blue light. Figs 6a and 6b show the coefficients of the first two eigenspectra labeled with the morphological type for the normalizations discussed in Section 4. Since the two eigenspectra define a plane, and each of the spectra are themselves normalized, the sum of the squared coefficients adds up to 1, within the 1% tolerance. Thus every spectrum can be uniquely defined by a mixing angle and the different spectra form a one parameter family.

The size of the points in Figs 6a and 6b are correlated with the relative error shown in Fig 5, for $n = 2$. By changing the normalization of the SED's from one using a scalar product to weighting by the morphological populations the uncertainty in describing the quiescent galaxies decreases. Our conclusion from these two examples is that, in order to describe the wide range of spectral types to 1% accuracy we are much better off using the scalar product normalization. If both the eigenspectra and the galaxies are normal vectors (in the 2207 dimensional space), their scalar product is just the cosine of the angle between them. On the other hand, by putting all the weight into the early type galaxies, they are better resolved with the two components, but the errors increase for the starburst galaxies.

The geometric interpretation is quite simple: the overall distribution of spectra can be described by a line on the surface of a multidimensional sphere. This line has a small degree of curvature but can be approximated quite well by a main circle, corresponding to the first two expansion coefficients. When we approximate a smaller subset of the distribution (e.g. the quiescent galaxies) with a different main circle, we effectively find the local tangent to the curved line, providing a more accurate local expansion. This can be used to improve on our simple technique, but more galaxy spectra have to be considered. We envisage an iterative scheme, where one would use two eigenspectra to perform a rough classification, i.e. to locate the region on the line, then use another set of two "local tangent" eigenfunctions to define the precise subtype, based upon the continuum SED.

We illustrate this point by including the third eigenspectrum in our classification. The variation of spectral type, as a function of expansion coefficients, is represented by spherical coordinates in Fig 6c. For the scalar product normalization the mean spectrum (or center of mass of the distribution) is defined by $\phi = 0$, $\theta = 0$. The main deviation from this spectrum (the second eigenspectrum) is given by the angle $\phi$, with spectra that are bluer or redder than the mean having positive and negative values respectively. The angle $\theta$ describes the contribution of the post starburst activity (third eigenspectrum) to the SED. There is a visible curvature within the spherical coordinate system with the S6 SED deviating by $8°$ from the equator. Even given this curvature the relation remains monotonic and can still be described in terms of one parameter.

The points shown in Fig 6c are derived using the scalar product normalization. We can, however, compare how well a particular classification scheme (i.e. number of eigen-components, $n$, and normalizations) describes the distribution of the data by transforming the expansion coefficients into this spherical coordinate system. For the scalar product norm and $n = 2$ the variation of spectral type is described by a line passing through the equator ($\theta = 0$). While this describes the overall distribution well it deviates by $\sim 10°$ at the center and extremes of the distribution. Normalizing by the morphological distribution (dotted line) moves the mean spectrum towards that of the quiescent galaxies ($\phi = -20$). The classification more accurately represents these quiescent galaxies at the expense of the starburst galaxies.

The next step in our iterative scheme is illustrated by dividing the distribution of spectral types into a red and blue population (split at $\phi = 0$). The dashed lines in Fig 6c show the classification derived from these samples independently. This piecewise linear representation provides the same accuracy as including a larger number of expansion coefficients. For our present data set and weighting schemes this simple two segment approximation provides the optimal classification scheme.

## 5. DISCUSSION

Sections 3 and 4 show that the continuum spectra of galaxies, ranging in morphological type from starburst to early type, can be described in terms of a two component, one parameter, family of spectra. If we consider the spectral shapes in terms of broadband photometry (treating the eigenspectra as the response function of an ideal set of filters) then we see that the combination of a UV filter and red filter (in the restframe of the galaxy) would be able to distinguish between the different spectral types of galaxies. This is consistent with the findings of Connolly *et al.* (1994) who show that galaxies lie in an almost planar distribution in the U, $B_J$, $R_F$, $I_N$ bands.

Other methods to spectrally classify galaxies on the basis of the shape of their spectral continuum have also found that a minimum of two components are needed. Aaronson (1978) found that by using A0 V and M0 III stars as basis spectra, one could form a one parameter family of spectra that described the U-V and V-K colors of galaxies of all Hubble types. With an independent sample, Bershady (1995) found that with the added spectral resolution of $B_J$ and $I_N$ bands, (in addition to U, V, and K) at least five basic families (five distinct stellar pairs) were needed to fit all the observed colors. Yet each of these families contained only two components, i.e. one parameter. This somewhat surprising result comes naturally out of our analysis in §4. If the distribution of spectra form a curve on the surface of a sphere then the general distribution can be approximated by a two component fit (i.e. the AO V and MO III stars). A better approximation is, however, found if we decompose the distribution into a set of line segments, i.e. the tangents to the curved line. Each of these line segments is again just a two component model but the parameters (or stars) for these lines change as we move along the distribution.

Our analysis shows that the distribution of spectral types separate naturally as a function of the mixing angle between the two eigenspectra. This result is independent of the morphological classifications derived from the visible appearance of the galaxies. We can, however, determine whether the spectral classifications are correlated with morphology. Each SED is the combination of a number of galaxy spectra. All spectra used in a particular spectral type have the same morphological type, as derived from the RSA (Sandage & Tammann 1987) and RC3 (de Vaucouleurs *et al.* 1991). We can, therefore, use the morphologies for the SED's for a comparison with the spectral classifier.

Fig 5 shows that there is a strong correlation between the spectral classification derived from the eigenspectra and that determined directly from the morphologies. This is not all together surprising as it is known that galaxy colors correlate with morphology. With our current analysis we can, however, discriminate not just between elliptical and spiral galaxies but also between subtypes. By adopting the hierarchical approach of §4.2, classification of galaxy spectra can be accomplished with a handful of template spectra. To derive the intrinsic resolution of our classification, due to the variation of spectral type within a given morphological type, we require a larger sample of high quality spectra.

Our findings can be compared with the PCA of QSO spectra undertaken by Francis *et al.* 1993. They find that the broad features of the QSO spectra can be accounted for by as few as three eigenspectra. However, to account for 90% of the power in the correlation matrix, they require the combination of over nine eigenspectra. A fundamental difference between the analysis presented here and that of Francis *et al.* is that we consider only the continuum spectra, removing the strong emission lines. Repeating the analysis including the emission lines does not change the conclusions of this paper as the equivalent widths of the emission lines are small. Inclusion of the strong emission lines can, however, lead to an overinterpretation of the higher order eigenspectra. This arises because emission lines of variable intensity and linewidth are not well suited to a linear orthogonal expansion. Higher order eigenspectra will predominantly concern the variation in linewidth and shape of the emission lines without necessarily containing much physical significance.

Our classification may be improved by including emission lines as a separate set of vectors, where the magnitude of each component corresponds to the equivalent width of the emission line. Further, by weighting the spectra in wavelength space we can emphasize different absorption features and, therefore, add metalicity as an additional axis in the classifications. These improvements will be addressed in a subsequent paper.

## 6. CONCLUSIONS

(1) The Karhunen-Loève transform provides an elegant method for the derivation of eigenfunctions that describe multidimensional datasets. When applied to galaxy spectra it provides a powerful tool for spectral classification and for the identification of systematic errors within data sets.

(2) From the analysis of the SED's the distribution of galaxy spectral types, from the UV through to the near-infrared, can be described by linear combinations of two eigenspectra.

(3) Decomposing the SED's into these eigenspectra their coefficients can be used as an objective spectral classification scheme. This natural classification can be described in terms of a one parameter problem and, in the mean, is strongly correlated with independent morphological classifications. We propose the adoption of this scheme for future spectroscopic surveys.

(4) The resolution of any classification can be optimized to a particular range of spectral types by adopting an iterative scheme where we vary the normalizations of the input SED's. Further refinements maybe possible using the emission line equivalent widths as additional parameters and weighting different spectral features (e.g. the MgI absorption line).

We thank Michael Vogeley, Gyula Szokoly and Robert Brunner for useful discussions during the development of this analysis. AJC acknowledges partial support from the NSF grant AST-9020380. AS acknowledges support from OTKA in Hungary, an NSF-Hungary Exchange Grant, the US-Hungarian Fund and the Seaver Foundation. MAB was supported by Hubble

Fig. 1.— The distribution of spectral types taken from Kinney *et al.* (1994). The morphologies of the galaxies used to construct these composite SED's were taken from the RSA and RC3 catalogs. All galaxy spectra are normalized by their scalar product.

Fig. 2.— Eigenspectra derived from the Kinney *et al.* (1994) SED's. Galaxy spectra were normalized by their scalar products before calculating the eigensystem. The eigenspectra are ordered by decreasing eigenvalue. From the fourth eigenmode the spectra are already describing the noise distribution of the data (random and systematic). Artifacts due to errors in the data reduction, i.e. residual sky lines, show up in the later eigenmodes.

Fig. 3.— The first two eigenspectra derived from the SED's, after correcting the norm for the observed distribution of spectral types. The first component contains over 99% of the power within the correlation matrix (see text). Eigenspectra beyond the first two are essentially noise.

Fig. 4.— Each SED is decomposed into a linear combination of eigenspectra. We reconstruct the SED's using just the first two components (the eigenspectra were normalized by their scalar products). Each figure shows the reconstructed SED together with the residuals from the original spectrum. Even using the first two components we can reconstruct the input data to a mean accuracy of 1%. The starburst spectra of S1 and S6 have the largest deviations as their input spectra show evidence for post starburst activity.

Fig. 5.— We derive the relative error between the reconstructed SED's (i.e. $n < m$) and the original data, as a function of spectral type. Fig 5a shows the distribution of relative errors for the eigenspectra derived using the scalar product normalization. The graphs are unshaded for $n = 2$ and and shaded for $n = 3$. Fig 5b shows the relative errors for the eigenspectra derived from the morphological normalization. The shading is as given previously.

Fig. 6.— We show the distribution of coefficients determined by decomposing the SED's into their eigenspectra. The variation in spectral type is seen to be a one parameter family (described by the mixing angle between the two eigenspectra). Fig 6a shows the correlation between the first two coefficients and the morphological types of the galaxies (for the normalization by the scalar product). The size of the points correlates with their relative errors, for $n = 2$ (see text). Fig 6b shows the correlation of morphology and spectral type when weighting by the observed distribution of morphological types. Note that the errors for the quiescent galaxies improve while those for the starburst galaxies increase. Including the coefficient from the third eigenspectrum we show the distribution of mixing angles in terms of their spherical coordinates, centered on the first eigenspectrum (Fig 6c). The solid line at $\theta = 0$ corresponds to the second eigenspectrum and the angle $\theta$ to the third eigenspectrum (i.e. the post starburst activity). There is a small curvature in the distribution of spectral types; even in this coordinate system the distribution of spectral types remains a one parameter family. The dotted line describes the two component eigensystem determined from the morphological weighting (the fiducial point $\times$ is the center of mass of the system). To illustrate our iterative classification scheme we split at the sample at $\phi = 0$ then derive the two eigensystems. The dashed lines represent the mixing of the first two eigenspectra in the respective samples.

Table 1.  Eigenvalues and Weights Normalized by the Scalar Product

| Eigenvalue | Power | S1 | S2 | S3 | S4 | S5 | S6 | Sb | Sa | S0 | E |
|---|---|---|---|---|---|---|---|---|---|---|---|
| 8.136 | 81.36 | 0.278 | 0.282 | 0.325 | 0.336 | 0.346 | 0.346 | 0.320 | 0.312 | 0.304 | 0.305 |
| 1.761 | 17.61 | 0.452 | 0.441 | 0.266 | 0.210 | 0.075 | -0.016 | -0.303 | -0.343 | -0.371 | -0.364 |
| 0.053 | 0.53 | -0.412 | -0.330 | 0.207 | 0.031 | 0.391 | 0.619 | -0.135 | -0.106 | -0.254 | -0.217 |
| 0.014 | 0.15 | 0.085 | 0.186 | -0.863 | 0.272 | 0.152 | 0.288 | -0.025 | -0.017 | -0.163 | 0.079 |
| 0.011 | 0.11 | -0.031 | 0.106 | -0.153 | -0.115 | 0.272 | -0.126 | 0.335 | 0.181 | 0.307 | -0.789 |
| 0.021 | 0.02 | 0.030 | 0.004 | 0.042 | -0.060 | 0.014 | -0.135 | 0.009 | 0.762 | -0.626 | -0.038 |
| 0.004 | 0.04 | 0.157 | -0.049 | 0.013 | -0.261 | -0.043 | 0.084 | 0.773 | -0.352 | -0.409 | 0.084 |
| 0.005 | 0.05 | -0.542 | 0.138 | 0.071 | 0.600 | 0.142 | -0.451 | 0.221 | -0.148 | -0.157 | 0.058 |
| 0.006 | 0.06 | -0.471 | 0.733 | 0.045 | -0.359 | -0.220 | 0.241 | -0.009 | 0.033 | -0.001 | 0.050 |
| 0.007 | 0.07 | -0.008 | -0.108 | 0.056 | 0.441 | -0.744 | 0.333 | 0.165 | 0.114 | 0.031 | -0.291 |

Table 2.  Eigenvalues and Weights Normalized by the Total Flux

| Eigenvalue | Power | S1 | S2 | S3 | S4 | S5 | S6 | Sb | Sa | S0 | E |
|---|---|---|---|---|---|---|---|---|---|---|---|
| 4.756E-03 | 79.46 | 0.338 | 0.335 | 0.323 | 0.322 | 0.314 | 0.309 | 0.305 | 0.305 | 0.305 | 0.304 |
| 1.171E-03 | 19.56 | 0.472 | 0.450 | 0.216 | 0.159 | 0.036 | -0.042 | -0.303 | -0.348 | -0.384 | -0.375 |
| 3.050E-05 | 0.50 | -0.397 | -0.292 | 0.302 | 0.098 | 0.407 | 0.596 | -0.115 | -0.097 | -0.254 | -0.220 |
| 8.141E-06 | 0.14 | 0.023 | 0.197 | -0.845 | 0.256 | 0.179 | 0.324 | -0.025 | -0.014 | -0.187 | 0.095 |
| 6.499E-06 | 0.11 | -0.081 | 0.157 | -0.156 | -0.091 | 0.192 | -0.074 | 0.328 | 0.199 | 0.326 | -0.800 |
| 4.368E-06 | 0.07 | 0.615 | -0.730 | -0.103 | 0.216 | 0.049 | 0.005 | 0.042 | 0.028 | 0.035 | -0.160 |
| 3.350E-06 | 0.06 | -0.149 | 0.018 | 0.076 | 0.513 | -0.756 | 0.268 | 0.147 | 0.091 | -0.021 | -0.187 |
| 2.521E-06 | 0.04 | 0.287 | 0.065 | -0.043 | -0.612 | -0.291 | 0.568 | -0.259 | 0.191 | 0.152 | -0.058 |
| 2.160E-06 | 0.04 | -0.118 | 0.002 | -0.018 | 0.316 | 0.082 | -0.131 | -0.776 | 0.311 | 0.398 | -0.067 |
| 9.486E-07 | 0.02 | 0.022 | 0.005 | 0.044 | -0.059 | 0.023 | -0.160 | 0.001 | 0.772 | -0.609 | -0.039 |

Table 3.   Eigenvalues and Weights Normalized to the Distribution of Morphological Types

| Eigenvalue | Power | S1 | S2 | S3 | S4 | S5 | S6 | Sb | Sa | S0 | E |
|---|---|---|---|---|---|---|---|---|---|---|---|
| 9.792 | 97.92 | 0.032 | 0.034 | 0.048 | 0.053 | 0.060 | 0.064 | 0.597 | 0.534 | 0.415 | 0.414 |
| 0.168 | 1.68 | 0.478 | 0.475 | 0.394 | 0.364 | 0.286 | 0.228 | 0.188 | -0.086 | -0.212 | -0.193 |
| 1.975E-02 | 0.20 | -0.123 | -0.110 | -0.079 | -0.090 | -0.040 | -0.046 | 0.416 | 0.102 | 0.180 | -0.859 |
| 1.055E-02 | 0.11 | 0.110 | 0.107 | 0.110 | 0.076 | 0.075 | 0.061 | -0.657 | 0.564 | 0.384 | -0.226 |
| 5.978E-03 | 0.06 | -0.278 | -0.230 | 0.003 | 0.009 | 0.173 | 0.343 | 0.017 | 0.506 | -0.682 | -0.033 |
| 1.931E-03 | 0.02 | 0.347 | 0.303 | -0.329 | -0.016 | -0.401 | -0.523 | 0.047 | 0.340 | -0.356 | -0.022 |
| 6.742E-04 | 0.01 | -0.052 | -0.167 | 0.843 | -0.246 | -0.286 | -0.317 | 0.022 | 0.077 | -0.097 | 0.007 |
| 2.252E-04 | 0.00 | -0.560 | 0.143 | 0.063 | 0.662 | 0.138 | -0.453 | -0.008 | 0.003 | -0.001 | 0.000 |
| 2.890E-04 | 0.00 | -0.477 | 0.724 | 0.038 | -0.318 | -0.264 | 0.277 | 0.002 | -0.008 | 0.013 | 0.005 |
| 3.413E-04 | 0.00 | 0.038 | -0.182 | 0.022 | 0.501 | -0.739 | 0.406 | 0.000 | -0.030 | 0.047 | -0.018 |

Table 4.   Expansion Coefficients of the Eigenspectra Normalized by the Scalar Product

| SED | e1 | e2 | e3 | e4 | e5 | e6 | e7 | e8 | e9 | e10 |
|---|---|---|---|---|---|---|---|---|---|---|
| S1 | 0.7927 | 0.5997 | -0.0955 | 0.0103 | -0.0032 | 0.0012 | 0.0097 | -0.0367 | -0.0358 | -0.0007 |
| S2 | 0.8044 | 0.5858 | -0.0766 | 0.0225 | 0.0111 | 0.0002 | -0.0030 | 0.0093 | 0.0558 | -0.0089 |
| S3 | 0.9283 | 0.3532 | 0.0481 | -0.1046 | -0.0160 | 0.0017 | 0.0008 | 0.0048 | 0.0034 | 0.0046 |
| S4 | 0.9575 | 0.2793 | 0.0071 | 0.0330 | -0.0120 | -0.0024 | -0.0161 | 0.0407 | -0.0273 | 0.0362 |
| S5 | 0.9882 | 0.0995 | 0.0907 | 0.0184 | 0.0284 | 0.0006 | -0.0026 | 0.0096 | -0.0167 | -0.0611 |
| S6 | 0.9877 | -0.0212 | 0.1435 | 0.0349 | -0.0131 | -0.0054 | 0.0052 | -0.0306 | 0.0183 | 0.0274 |
| Sb | 0.9127 | -0.4026 | -0.0313 | -0.0030 | 0.0351 | 0.0004 | 0.0477 | 0.0149 | -0.0007 | 0.0135 |
| Sa | 0.8891 | -0.4550 | -0.0247 | -0.0021 | 0.0189 | 0.0305 | -0.0217 | -0.0100 | 0.0025 | 0.0094 |
| S0 | 0.8666 | -0.4927 | -0.0589 | -0.0197 | 0.0320 | -0.0250 | -0.0252 | -0.0107 | 0.0000 | 0.0026 |
| E | 0.8696 | -0.4835 | -0.0504 | 0.0095 | -0.0824 | -0.0015 | 0.0052 | 0.0039 | 0.0038 | -0.0239 |

Table 5.  Expansion Coefficients of the Eigenspectra Normalized by the Distribution of
Morphological Types

| SED | e1 | e2 | e3 | e4 | e5 | e6 | e7 | e8 | e9 | e10 |
|-----|------|------|------|------|------|------|------|------|------|------|
| S1 | 0.4539 | 0.8768 | -0.0774 | 0.0506 | -0.0962 | 0.0683 | -0.0061 | -0.0376 | -0.0363 | 0.0031 |
| S2 | 0.4700 | 0.8708 | -0.0693 | 0.0491 | -0.0795 | 0.0596 | -0.0194 | 0.0096 | 0.0551 | -0.0150 |
| S3 | 0.6785 | 0.7217 | -0.0495 | 0.0506 | 0.0009 | -0.0646 | 0.0979 | 0.0042 | 0.0029 | 0.0018 |
| S4 | 0.7381 | 0.6676 | -0.0569 | 0.0350 | 0.0032 | -0.0032 | -0.0285 | 0.0445 | -0.0242 | 0.0414 |
| S5 | 0.8414 | 0.5246 | -0.0248 | 0.0346 | 0.0599 | -0.0787 | -0.0332 | 0.0093 | -0.0200 | -0.0611 |
| S6 | 0.8916 | 0.4183 | -0.0289 | 0.0281 | 0.1186 | -0.1028 | -0.0368 | -0.0304 | 0.0211 | 0.0335 |
| Sb | 0.9980 | 0.0412 | 0.0313 | -0.0361 | 0.0007 | 0.0011 | 0.0003 | -0.0001 | 0.0000 | 0.0000 |
| Sa | 0.9988 | -0.0211 | 0.0085 | 0.0346 | 0.0234 | 0.0089 | 0.0012 | 0.0000 | -0.0001 | -0.0003 |
| S0 | 0.9962 | -0.0665 | 0.0194 | 0.0303 | -0.0404 | -0.0120 | -0.0019 | 0.0000 | 0.0002 | 0.0007 |
| E | 0.9937 | -0.0606 | -0.0927 | -0.0178 | -0.0019 | -0.0007 | 0.0001 | 0.0000 | 0.0001 | -0.0003 |

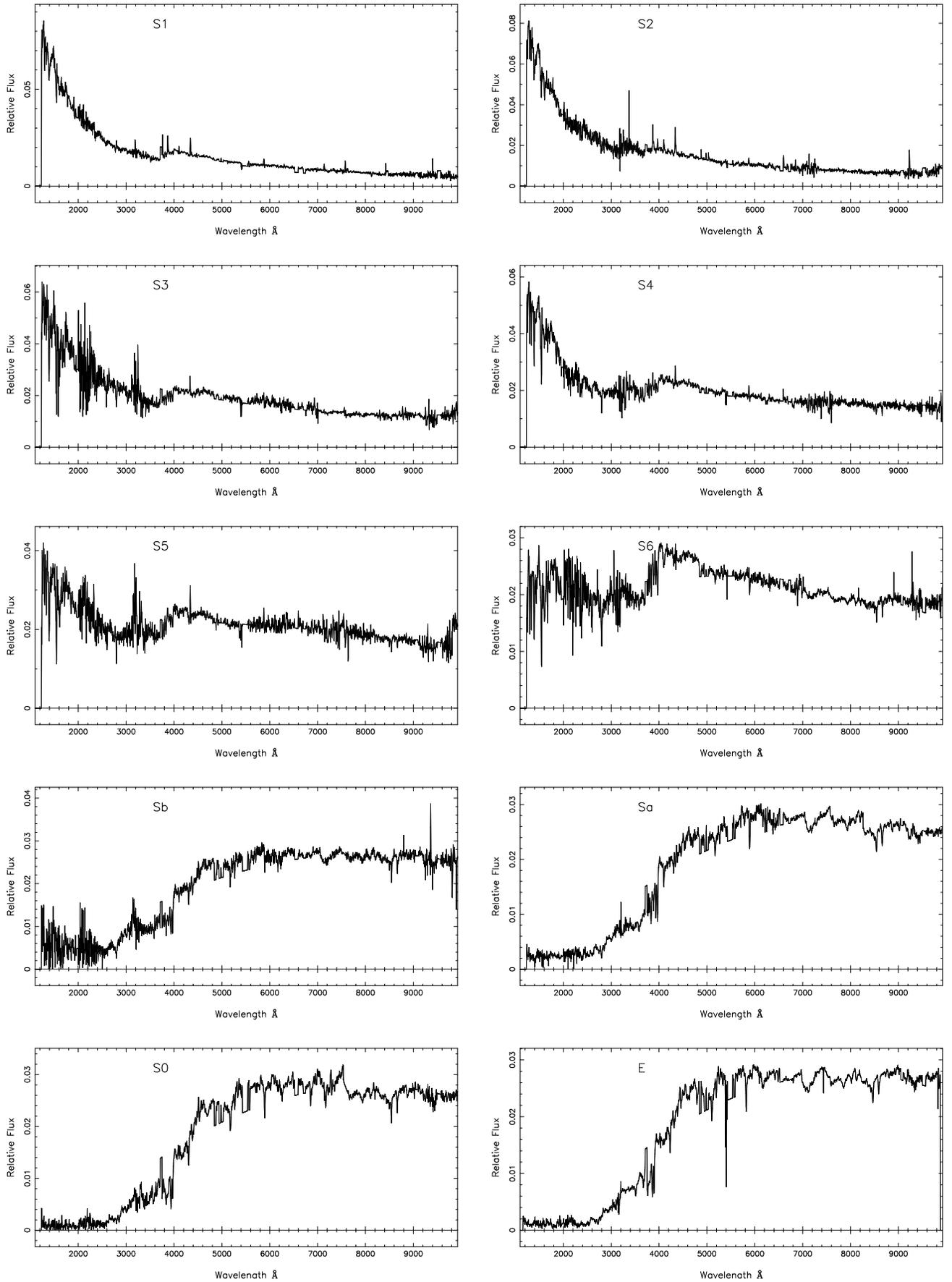

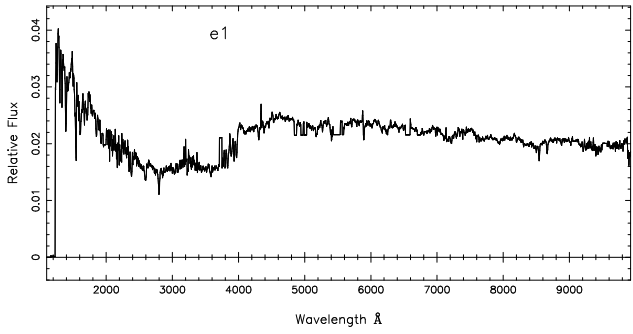

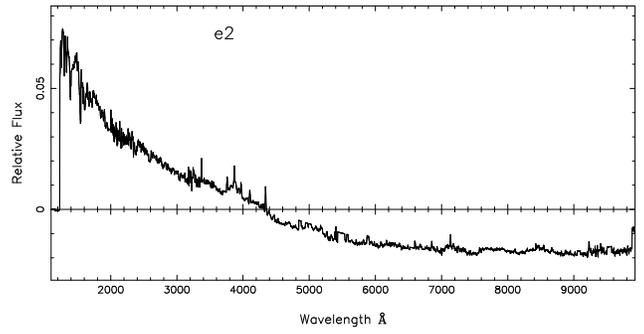

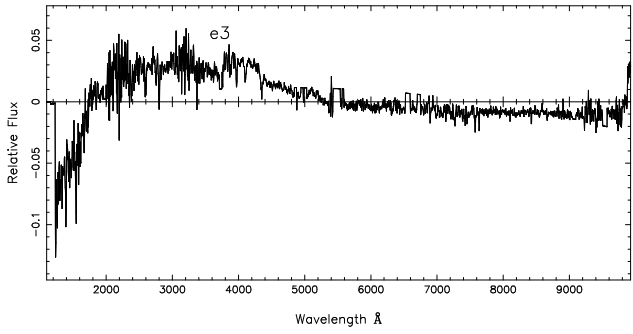

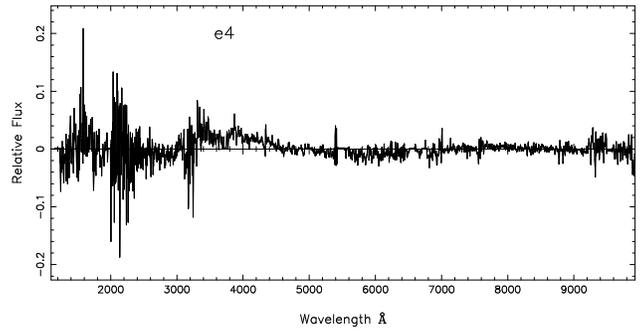

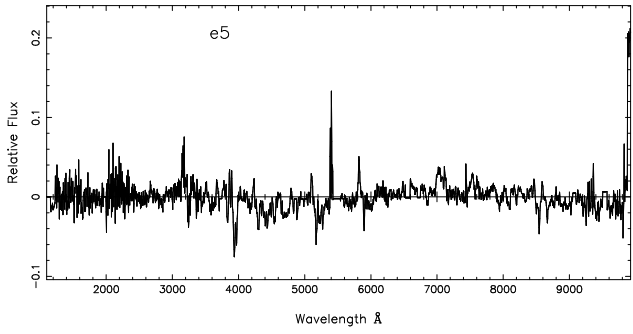

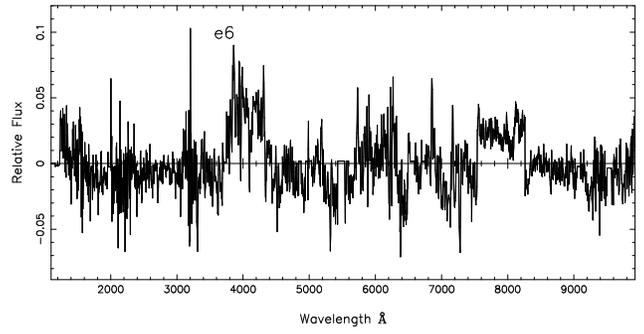

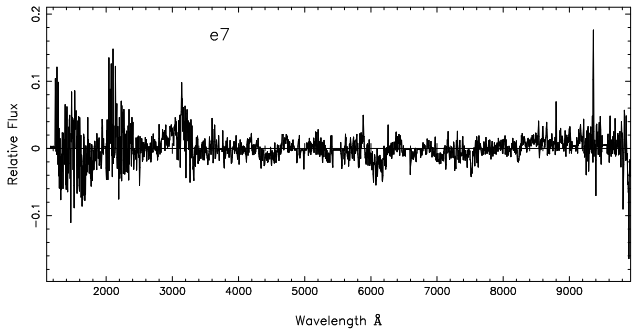

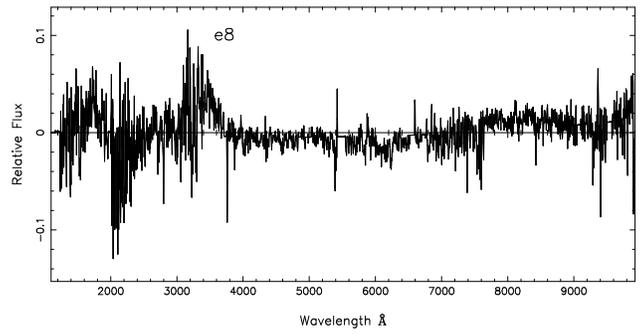

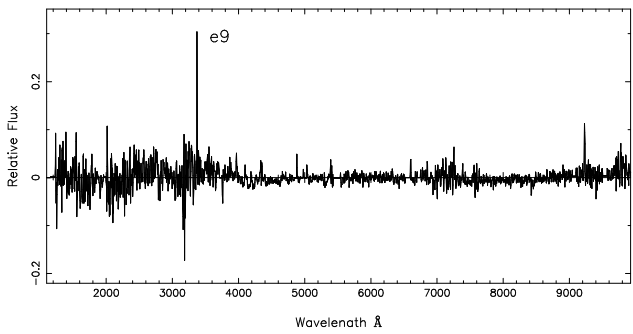

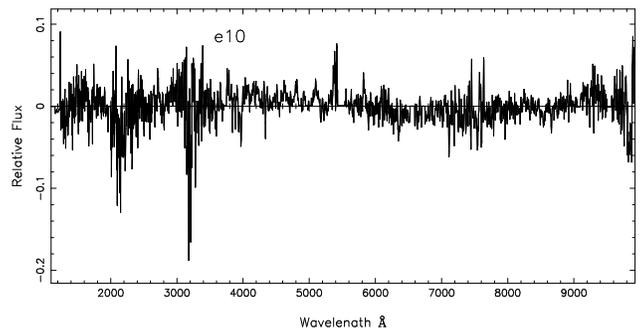

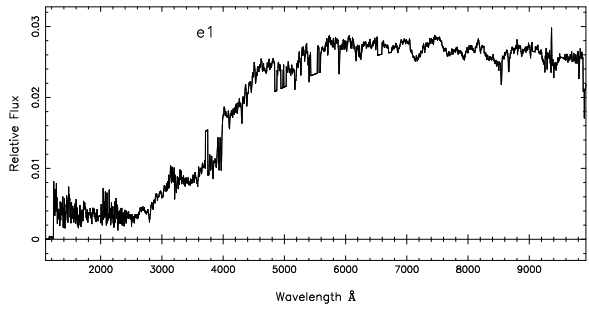
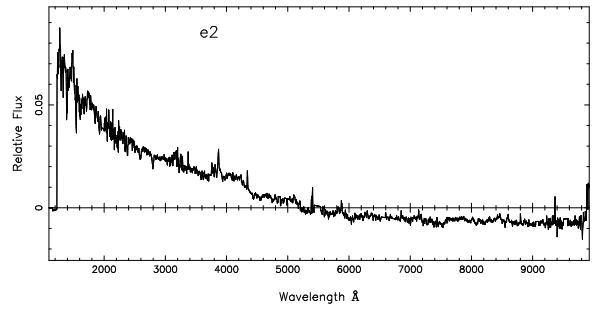

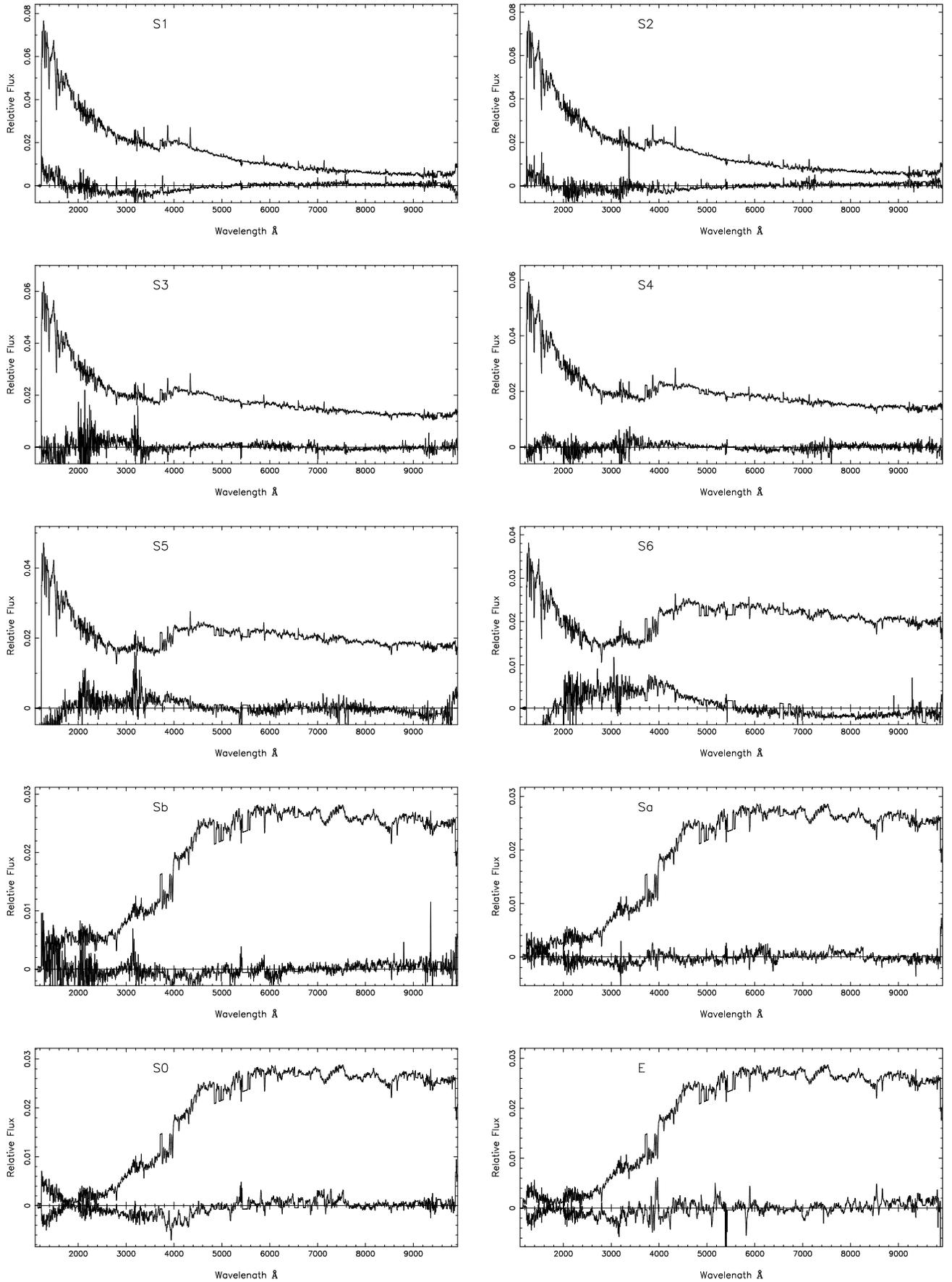

Figure 5

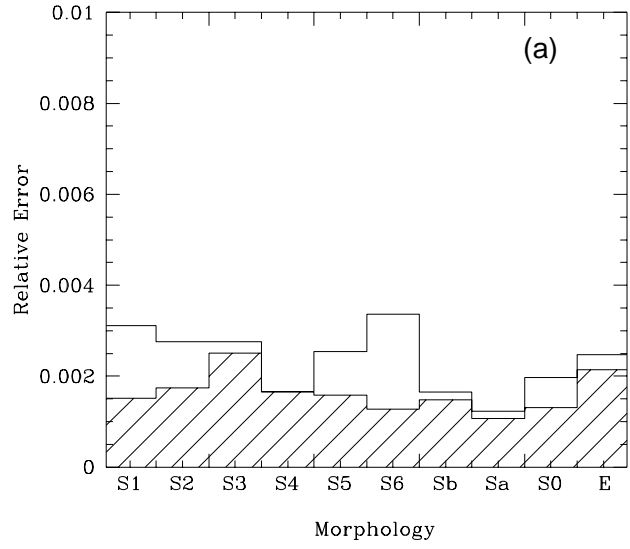

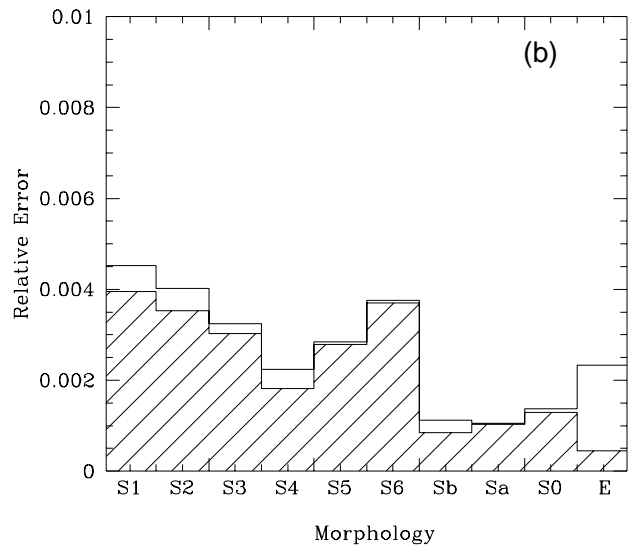

Figure 6

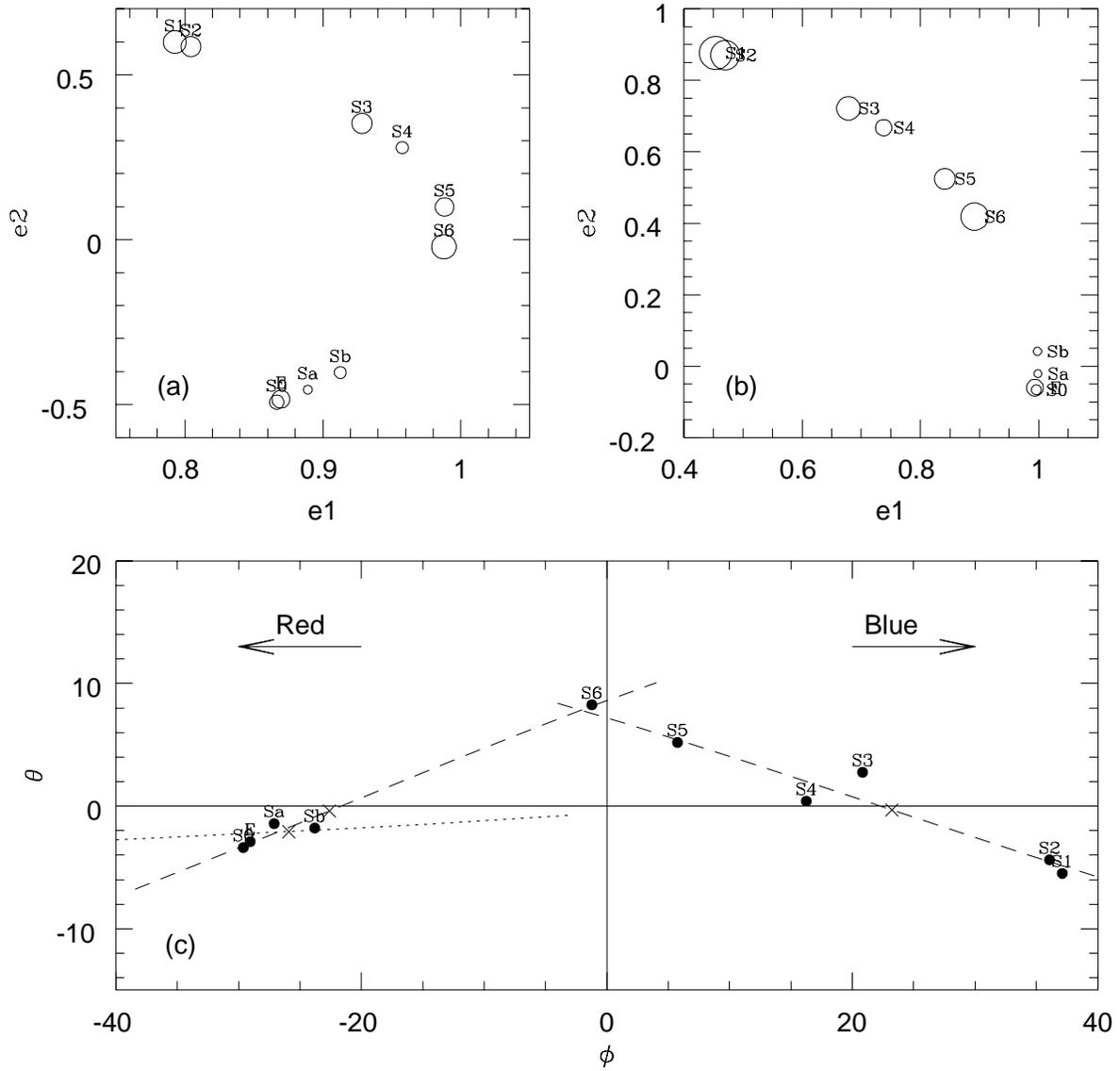